\documentclass[aps,prd,twocolumn,showpacs,groupedaddress,floatfix]{revtex4}

\usepackage{graphicx}
\usepackage{longtable}
\DeclareGraphicsExtensions{.pdf, .eps, .jpg, .png}

\begin{document}

\title{Limits of dark energy measurements from CMB lensing-ISW-galaxy count correlations}
\author{Benjamin Gold}
\email{gold@bubba.ucdavis.edu}
\affiliation{Department of Physics, UC Davis,
One Shields Avenue, Davis CA, 95616}

\date{\today}

\begin{abstract}
I discuss several issues that arise when trying to constrain the dark energy equation
of state using correlations of the integrated Sachs-Wolfe effect with galaxy counts and lensing of the cosmic microwave background. 
These techniques are complementary to others such as galaxy shear surveys, and can use data that will already be obtained from currently planned observations.
In regimes where cosmic variance and shot noise are the dominant sources of error,
constraints could be made on the mean equation of state to $\pm 0.33$ and its first derivative to $\pm 1.0$.
Perhaps more interesting is that the determination of dark energy parameters by these types of experiments depends strongly on the presence or absence of perturbations in the dark energy fluid. 

\end{abstract}


\maketitle

\section{Introduction}
Understanding cosmic acceleration is one of the largest problems facing physics today.  So far the most direct measurements of acceleration have come from distance-redshift measurements \cite{Perlmutter:1998np, Riess:1998cb, Knop:2003iy}.  This acceleration is thought to be due to the effect of dark energy, a new form of energy density that dominates the current universe.  However, a universe with dark energy exhibits a different evolution for density perturbations, and this has observable consequences.

In this paper, I will discuss how well certain measurements of growth can constrain the expansion history of the universe, and thus dark energy.  In particular I will discuss the lensing of the cosmic microwave background (CMB) by intervening structure combined with measurements of galaxy number density correlations from surveys.  These types of measurements are attractive since they can be obtained essentially ``for free'' from experiments already planned.  I will also discuss the sensitivity of these measurements to the presence of perturbations in the dark energy fluid, which can have an important effect on the results \cite{Caldwell:1997ii}.
In this paper ``lensing'' refers specifically to lensing of the CMB by foreground structure, rather than the extremely productive research area of studying the shape distortion of background galaxies due to lensing from foreground galaxies (covered extensively in \cite{Jain:2003tb, Song:2003gg}).


\section{Background theory}
\subsection{The ISW effect and dark energy} 
As photons travel from the last scattering surface (LSS) to us, they fall into and climb out of potential wells that lie along their path.  If the gravitational potential $\Phi$ does not change with time, then the accompanying blueshifts and redshifts will cancel each other out, leaving no net effect.  However, if the potential does change over time there may be some overall change in each photon's wavelength, and hence the observed temperature.  This is the integrated Sachs-Wolfe (ISW) effect, with the change in temperature $\Delta T^\mathrm{ISW}$ observed in a direction $\hat\mathbf{n}$ expressed simply as
\begin{equation}
	\Delta T^\mathrm{ISW} (\hat\mathbf{n}) = -2 \int dD\, \dot \Phi(x(\hat\mathbf{n}),D)   ,
\end{equation}
where the gravitational potential is written as a function of position $x$ and lookback distance $D$ (used as a proxy for conformal time).

	
For a flat, matter-dominated universe, the potential remains constant over time even though the density perturbations themselves do not.  This means that any ISW effect originates either from early times, when the density of radiation was still significant enough to affect the expansion rate, or from late times when dark energy became dominant.  Thus if the late ISW signal can be separated out from others, it provides a clean measurement of dark energy.

Normally the ISW effect itself is buried by the primary temperature anisotropies from the LSS.
However, the primary anisotropies are set at the last scattering surface, at a completely different epoch and at different length scales than those of the structure growth responsible for the ISW effect.  The primary anisotropy should thereby be uncorrelated with the ISW effect and other measurements of growth.
Thus cross-correlating other measurements with CMB temperature maps can be a useful tool for bringing out information about growth, as has been discussed recently by \cite{Afshordi:2004kz, Peiris:2000kb} and demonstrated by \cite{Scranton:2003in}.

\subsection{Lensing correlations}

The first measurement I will consider correlating with the temperature map is a measurement of gravitational lensing of the CMB.  The microwave background is gravitationally lensed by matter that lies between us and the LSS.  The map of photon deflection angles over the sky can be written as the gradient of a scalar field $\phi$ called the projected potential, which depends on the 3D gravitational potential $\Phi$ as
\begin{equation}
	\phi(\hat \mathbf{n}) = -2 \int dD\, \frac{D_\mathrm{LSS} - D}{D\, D_\mathrm{LSS}} \Phi \left(
	\mathbf{x}(\hat\mathbf{n}), D \right)   ,
\end{equation}
where $D_\mathrm{LSS}$ is the distance to the last scattering surface.

Just as with CMB temperature maps, the map of the projected potential can be decomposed into spherical harmonics.  Then a two-point function of the modes can be put together to construct an angular power spectrum.  The expression for the power spectrum can be written as a line-of-sight integral over the 3D gravitational potential 
\begin{equation}
	C_\ell^{\phi\phi} \sim \int \left. dD \, D \, \Phi^2(k,D) \left( \frac{D_S-D}{D\,D_S} \right)^2
	P^2_\delta(k) \right|_{k=\ell \frac{H_0}D}   .
\end{equation}
While this equation is only exact in the flat-sky (large $\ell$) limit, the important feature is that it captures the physics of how the angular power spectrum depends on a line-of-sight integral of the gravitational potential multiplied by a kind of window function and the \emph{primordial} (i.e. the growth function has been separated out) power spectrum $P^2_\delta(k)$.  The exact expressions and further details can be found in \cite{Hu:2000ee}; they are used for the computations in this paper for multipole moments where the difference is important.  However, for the remainder of this section I will write only the Limber-approximated integrals for clarity.  The cross correlation between the temperature and lensing is due to the presence of the ISW effect in the temperature, and thus its power spectrum has the simple flat-sky form
\begin{equation}
	C_\ell^{T\phi} \sim \int \left. dD \, D \, \Phi(k,D) \dot\Phi(k,D) \frac{D_S-D}{D\,D_S} P^2_\delta(k)
	\right|_{k=\ell \frac{H_0}D}  .
\end{equation}

\begin{figure}
	\includegraphics[width=3.4in]{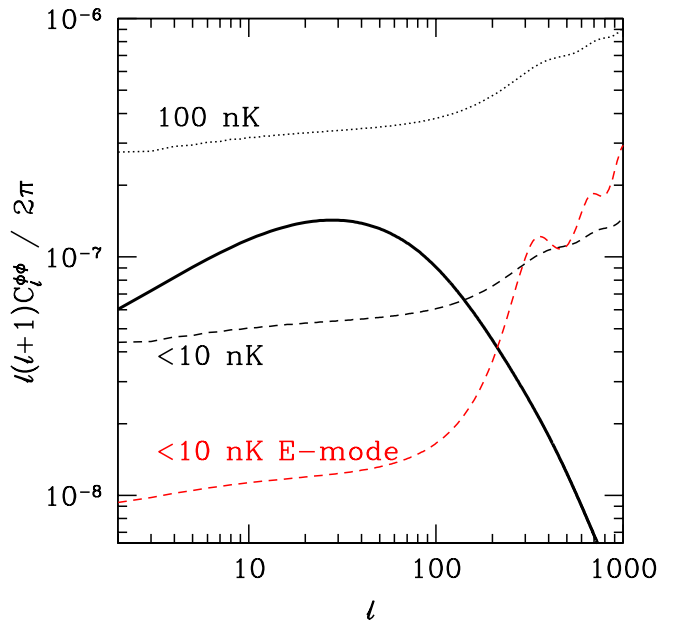}
	\caption{
		\label{fig:lensnoise}
		Noise levels for CMB lensing reconstruction.
		The dark curve is the lensing power spectrum for a typical cosmology.
		The different dotted lines are noise levels
		for CMB temperature experiments with a $4'$ beam and $0.1\,\mu K$ noise per pixel
		(dotted line) or $0.01\,\mu K$ noise per pixel or less (dashed line).  The lower dashed line is
		for an E-mode CMB polarization measurement.
	}
\end{figure}

\subsection{Galaxy correlations}
The second type of observation I will consider is counting galaxies projected on the sky.  On large scales, fluctuations in the number density of galaxies should track fluctuations in the gravitational potential (possibly with some bias).  From the map of number density over the sky, we can obtain the auto-correlation power spectrum
\begin{equation}
	C_\ell^{gg} \sim \int \left. dD \, D \, \Phi^2(k,D) n_g^2(D) P_\delta^2(k) \right|_{k=\ell\frac{H_0}D}
	  ,
\end{equation}
which depends on the potential as viewed by some window function $n_g(D)$  describing the distribution of galaxies that are actually observed.
The galaxy spectrum should be correlated both with the ISW part of the temperature spectrum
\begin{equation}
	C_\ell^{Tg} \sim \int \left. dD \, D \, \Phi(k,D)\dot\Phi(k,D) n_g(D) P_\delta^2(k) 
	\right|_{k=\ell\frac{H_0}D}   ,
\end{equation}
and with the lensing potential power spectrum
\begin{equation}
	C_\ell^{g\phi} \sim \int \left. dD \, D \, \Phi^2(k,D) n_g(D) \frac{D_S-D}{D\,D_S}
	P_\delta^2(k) \right|_{k=\ell\frac{H_0}D}   .
\end{equation}
Again, in the calculations described later these Limber-approximated integrals are used only for high multipole moments.  The exact expressions  are used for low multipole moments ($\ell < 100$).

\section{Calculating the sensitivity to dark energy}
The power spectra themselves are computed numerically using the techniques described in the previous section and a version of the \textsc{cmbfast} code \cite{cmbfast} which I modified to output lensing spectra and other information.

The basis for all the analysis is the Fisher information matrix.  Given a fiducial model, the Fisher matrix describes how sensitive the model is to changes in its parameters.  First, all the power spectra are put together into a covariance matrix
\begin{equation}
	\mathbf{C}_\ell \equiv \left( \begin{array}{ccc}
		C_\ell^{TT} + N_\ell^T & C_\ell^{T\phi} & C_\ell^{Tg} \\
		C_\ell^{T\phi} & C_\ell^{\phi\phi} + N_\ell^\phi & C_\ell^{\phi g} \\
		C_\ell^{Tg} & C_\ell^{\phi g} & C_\ell^{gg} + N_\ell^g\\
	\end{array}
	\right)   .
\end{equation}
Experimental noise can be modeled as a contribution to the diagonal terms of this matrix.  In our case I take the noise contribution to the primary temperature anisotropy $N_\ell^T$ to be negligible.  However, even a small amount of noise is significant when reconstructing lensing information, and I take this into account using the derivations for lensing error as a function of temperature noise found in \cite{Okamoto:2003zw} and shown in Fig.~\ref{fig:lensnoise}.  I take the primary contribution to the noise for galaxy surveys $N_\ell^g$ to be shot noise (bias for each survey is treated as a free model parameter).  Some other possible noise sources are discussed in section \ref{sec:gal}.

When I consider combining three galaxy surveys at different redshifts, each galaxy survey gets its own row and column, resulting in a $5\times5$ covariance matrix.  The Fisher matrix is then constructed as a sum over multipole moments,
\begin{equation}
	\label{eq:fish}
	F_{ij} \equiv \sum_\ell \textrm{Tr} \left[ \mathbf{C}_\ell^{-1} \frac{\partial \mathbf{C}_\ell}{\partial s_i}
		\mathbf{C}_\ell^{-1} \frac{\partial \mathbf{C}_\ell}{\partial s_j} \right]   ,
\end{equation}
where the $s_i$ are labels for the actual model parameters.

The model parameters actually varied are the dark energy parameters discussed in the next section, plus an angular scale parameter $\ell_A$, baryon density $\Omega_b h^2$, matter density $\Omega_m h^2$, primordial power amplitude $A_S$ and tilt $n_S$, optical depth $\tau$, and a bias parameter $b_i$ for each galaxy survey.  All models were constrained to be flat.
I investigated two fiducial models; both had parameters equivalent to a modern  concordance cosmology ($\Omega_\mathrm{d.e.} = 0.75$, $\Omega_b = 0.04$, $\Omega_\mathrm{cdm} = 0.21$, $h=0.65$, $\tau = 0.1$, $n_S = 0.9$).  The difference between the two was in the dark energy evolution; one model had a ``pure $\Lambda$'' equation of state of  $w = -1$, whereas the other used a quintessence-type model with $w=-0.9$ today running smoothly to $w = -1$ at high redshift.  The latter has a somewhat enhanced sensitivity to dark energy parameters due to the evolution of dark energy at low redshifts, and is used for reporting the final constraints on dark energy parameters in section \ref{sec:results}.  Using the ``pure $\Lambda$'' model instead degrades the dark energy parameter constraints by about 20\% overall.


\subsection{Choosing a form for dark energy}
Different experiments have differing sensitivities to dark energy.  With no knowledge of the fundamental physics behind cosmic acceleration, there is little reason to favor one functional form for the equation of state $w(z)$ over another.  In that light, principal mode analysis is useful for revealing what the experiments can actually constrain.  In such an analysis one chooses some basis set of eigenfunctions (cut off to a finite set size), and then expresses the fundamental modes for each experiment as combinations of basis functions.  In our case the basis functions will simply be a set of 25 boxcar functions covering the redshift range from $z=0$ to $z=5$.

\begin{figure}
	\includegraphics[width=3.4in]{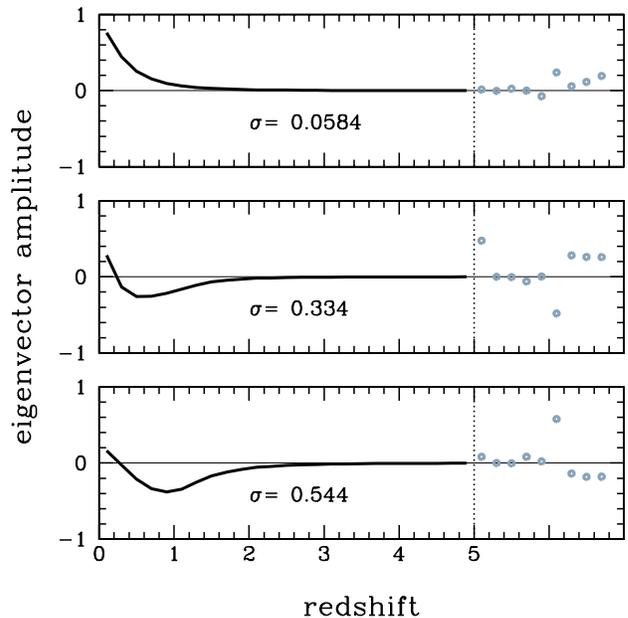}
	\caption{
		\label{fig:evec}
		Three dark energy eigenvectors for an ideal ISW-lensing-galaxy count correlation experiment.
		The left part of each plot shows the 25 $w(z)$ parameters, and on the right are the 9 other
		parameters (in order): angular scale $\ell_A$, baryon density $\Omega_b h^2$, matter
		density $\Omega_m h^2$, optical depth $\tau$, primordial power spectrum tilt $n_S$,
		primordial power amplitude $A_S$, and the bias parameters for each galaxy survey
		$b_1,b_2,b_3$.  The error in each eigenmode for a cosmic-variance limited experiment
		is shown for each mode.
	}
\end{figure}	
\begin{figure}
	\includegraphics[width=3.4in]{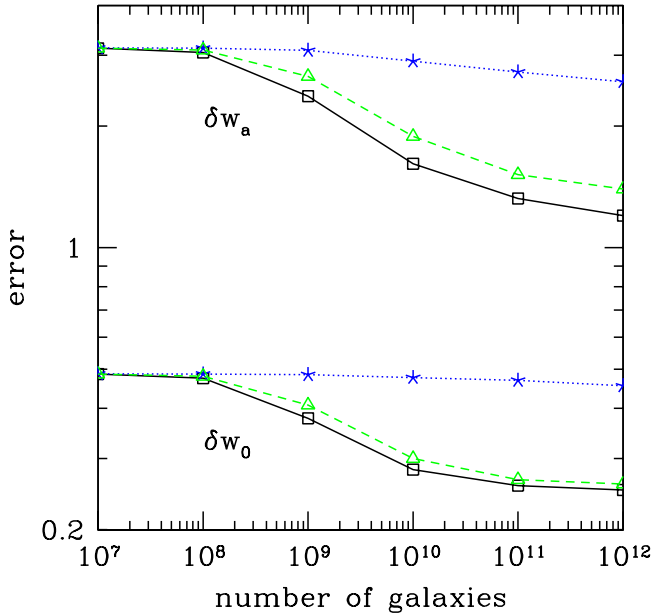}
	\caption{
		\label{fig:gplot}
		Error in $w_0$ and $w_a$ as a function of total galaxy number
		with lensing noise fixed at the ``$<10$ nK E-mode'' level from Fig.~\ref{fig:lensnoise}.
		The solid lines with squares are for a survey with three redshift bins,
		dashed lines with triangles  for a survey with one high redshift bin,
		and dotted lines with stars for one with a single low redshift bin.
	}
\end{figure}

A few example eigenvectors are shown in Fig.~\ref{fig:evec}.  In general, dark energy information is mixed with other cosmological parameters so that the eigenmodes are complicated combinations of several parameters.  However, 
with the Fisher matrix in hand not only can one perform principal mode analysis, one can also marginalize over some parameters and switch to more familiar dark energy parameters with simple matrix operations.  For example, it is simple to switch from the 25 $w(z)$ parameters to
 the ``$w_a$'' parametrization \cite{Chevallier:2000qy, Linder:2002et}, which takes the form
\begin{equation}
	w(z) = w_0 + w_a \frac{z}{1+z}
	  .
\end{equation}
The full eigenmode analysis shows that even for ideal experiments, no more than two or three dark energy parameters will be well-constrained, so for the rest of this paper the ``$w_a$'' parameterization will be assumed.

\subsection{Galaxy survey characteristics\label{sec:gal}}
My intent is to approximate the behavior of galaxy surveys which do not include spectra, but are able to obtain approximate photometric redshifts through color information.  I therefore consider the situation where one has a large number of galaxies that can be assigned to one of three redshift bins, centered around $z=0.5$, $z=1.0$, and $z=1.5$.  Each of these bins is approximated as a rounded boxcar function similar to the technique of Hu and Scranton \cite{Hu:2004yd}.  This approximation is best for survey slices that are not severely magnitude or volume limited.  For those cases the overall effect is to smear out the edges of the bins, weakening the advantage of having several redshift bins.

Galaxy surveys may contain numerous systematics.  Stellar contamination should not be a big problem since for faint surveys the majority of observed objects are galaxies anyway.
More worrisome is that the analysis presumes knowledge of the redshift distribution of galaxies in the sample, which of course could be in error.  We can estimate how small the overall redshift calibration error needs to be by converting the redshift error into an error in the angular scale of galaxy correlations, and examining the change in the correlation power spectrum compared to the shot noise.  Because of the large angular scales we're interested in, it turns out that even for galaxy surveys with $10^{12}$ galaxies (corresponding to roughly $10^4$ per square arcminute), shot noise at these scales is still large enough that an overall redshift calibration error of up to a percent is tolerable.  For the remainder of this paper I will assume any systematics in the galaxy correlations are below the shot noise level, but new experiments may well reveal important new systematics not considered here.


\section{Results\label{sec:results}}

In the absence of noise, over 40\% of the improvement in dark energy constraints gained by adding CMB lensing data to the temperature information comes from the $C_\ell^{T\phi}$ cross-correlation channel.
Even though in the ideal case the noise in the CMB and lensing potential  anisotropies is uncorrelated, the inversion of the covariance matrix in equation \ref{eq:fish} means that noise in the auto-correlation spectra will find its way into the cross-correlation channel.  This then means that the noise in the reconstructed $C_\ell^{\phi\phi}$ spectrum will be the limiting factor as to how well CMB lensing can constrain dark energy.


In Fig.~\ref{fig:lensnoise} I have calculated the noise in the reconstructed $C_\ell^{\phi\phi}$ lensing power spectrum for several different types of experiments.  At small enough temperature noise levels (a bit above $10\,nK$ per pixel for $4'$ pixels) all of the noise in the lensing $C_\ell^{\phi\phi}$ is in fact coming from cosmic variance in the temperature $C_\ell^{TT}$.  Note that even at this limit, the noise in the lensing spectrum is still roughly the same order of magnitude as the signal.
Part of the low signal-to-noise ratio appears to be related to the choice of a red tilt ($n_S < 1$) in the primordial spectrum of the fiducial model.  This reduces CMB power at the small scales from which the lensing signal is reconstructed, leading to more relative noise in the lensing power spectrum.
The total noise in the lensing power spectrum can also be reduced further by combining several polarization modes as per \cite{Okamoto:2003zw}, which results in an improvement by a factor of a few, so it may not be necessary to drive the detector temperature all the way down to $10 nK$ to get the desired precision.  In any case cosmic variance is the ultimate limit that I will consider here, and we shall see that it is a fairly restrictive one, at least as far as $w(z)$ constraints are concerned.

\begin{figure}[t]
	\includegraphics[width=3.4in]{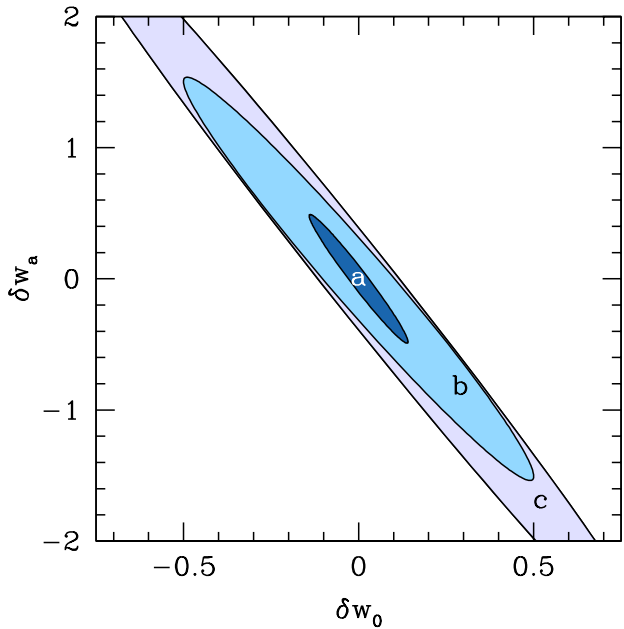}
	\caption{
		\label{fig:contours}
		Error contours (68.3\%) in the $w_0$--$w_a$ plane.
		From innermost to outermost they represent (a) 
		a ``perfect''
		CMB lensing-galaxy count cross correlation measurement where cosmic variance is
		the only limitation,
		(b) a more realistic measurement with galaxy density
		of  $\sim 10^2\, \textrm{arcmin}^{-2}$ and the ``$<10$ nK E-mode'' lensing noise of
		Fig.~\ref{fig:lensnoise},
		(c) an ISW-lensing only experiment (no galaxy counts).
	}
\end{figure}

How about cross-correlating galaxies with the ISW effect in order to augment our information about dark energy?  This can provide interesting results, especially with the possibility of adding even very limited redshift information into the mix.  I show the results in Fig.~\ref{fig:gplot}.
Galaxy information, especially that coming from high redshift galaxies, can tighten the dark energy constraints by more than a factor of two.

Finally, the constraint contours in the $w_0$--$w_a$ plane are shown in Fig.~\ref{fig:contours}.  These are constraints made on $w(z)$ simultaneously with constraints on all the other cosmological parameters described above.  In the absence of any other experiments, CMB lensing and galaxy counts combined can realistically constrain (contour b) $w_0$ to a precision of about $\pm 0.33$ when all other parameters are marginalized over.

\subsection{Dark energy perturbations}
\begin{figure}[t]
	\includegraphics[width=3.4in]{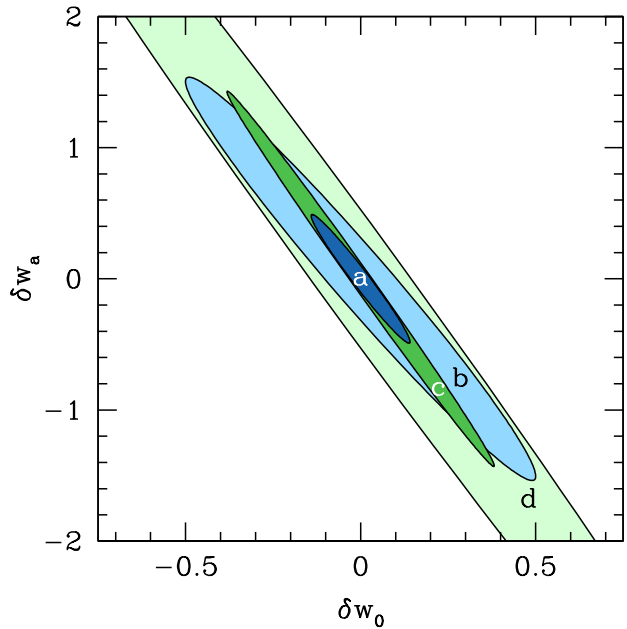}
	\caption{
		\label{fig:contoursp}
		The effect of dark energy perturbations on error contours (68.3\%) in the $w_0$--$w_a$ plane.
		Contours (a) and (b) are as before in Fig.~\ref{fig:contours}.  Contours (c) and (d) show
		the effect of including dark energy perturbations for experiments considered in (a) and (b),
		respectively.
	}
\end{figure}

The above results were obtained assuming the absence of perturbations in the dark energy fluid.  A true cosmological constant has $w=-1$ and is perfectly smooth with no perturbations.  However, many quintessence models are built with some sort of scalar field which in general can have its own density fluctuations.  These density fluctuations become important at late times when the structure formation responsible for the ISW effect is occurring, and thus significantly affect the above results.  This effect is shown in Fig.~\ref{fig:contoursp}.  
Note that including perturbations mostly increases the uncertainties along the degeneracy direction, especially for experiments limited only by cosmic variance, with the result that limits on both dark energy parameters are degraded by nearly a factor of three.

\section{Conclusion\label{sec:conclusion}}

Cross-correlating the ISW effect with CMB lensing and galaxy counts can in principle place limits on dark energy; ideally (if cosmic variance were the only limitation) one could measure $w_0$ to $\pm 0.093$ and $w_a$ to $\pm 0.32$, though in practice realistic limits are probably worse by about a factor of three.  These measurements depend on dark energy's effect on the growth of structure, and thus have a dependence on the equation of state quite different from distance-redshift measurements.  Observation of such growth effects would bolster the case for dark energy as the source of cosmic acceleration.  Also, these measurements are strongly sensitive to perturbations in the dark energy fluid (which do not affect distance observations) and thus may ultimately be more useful as a measurement of perturbations than as a precise determination of the equation of state.

There is certainly room for improvement for these kinds of measurements.  The amount of lensing being measured is small, so cosmic variance and the finite resolution of experiments become important.  In this work I have used the CMB temperature map to reconstruct the lensing spectrum; this method essentially makes use of four-point correlations in the original temperature map.  The three-point correlation function (known as the bispectrum) also contains contributions from lensing which are known to be significant \cite{Giovi:2003ri} and potentially less noisy, although how well dark energy information can be extracted from realistic data is still under investigation.

The scales on which galaxy densities correlate with the ISW effect are very large (degree scale or larger), so the deviations from a smooth background are small and even for large numbers of galaxies shot noise is significant.  One way of overcoming this problem is simply to use more information from galaxies than merely their number density.  Dedicated telescopes that can find galaxy shapes and redshifts are predicted to put good constraints on dark energy \cite{Song:2003gg}.
In the future a wide array of complementary observations will be available to determine the nature of cosmic acceleration, each with their own sensitivity and limitations.

\section*{Acknowledgments}
I would like to thank Andreas Albrecht and Lloyd Knox for providing much useful advice during the completion of this research.  I would also like to thank Manoj Kaplinghat and Mario Santos for productive discussions, and the anonymous referee for some extremely helpful corrections.  I was partially supported in this research by DOE grant 
DE-FG03-91ER40674.

\bibliography{darkmodes.bib}

\end{document}